# Sacred Landscape of the Incas in Central Chile

# Radial Ceque System Encoded in the Archeological Site Ruinas de Chada

**Nicolás Palacios-Prado and Fabiola Corominas**

**Abstract**

Mounts and hills played a predominant role in all pre-Hispanic Andean cultures, especially for the Inca culture. Through the use of georeferenced orthophotography, we found that the Inca site, Ruinas de Chada, represents the origin of a radial ceque system with alignments connecting; at one end high peaks of mounts of the Andes, and on the other end the summit of small hills in which important shrines were built. These alignments extend over two hundred kilometers, thus we propose that the information codified on this shrine was based on an ancient geodetic science. A sacred geometric relationship is encrypted in the pattern formed by the shrine´s positions with high accuracy, in which the Andean Chakana symbol is represented. These findings suggest that the valley of Chada could have been the sacred center of Collasuyu.

**Introduction**

There are records of innumerable ritual sites and mountain shrines throughout the Inca empire (*1*), reflecting an important devotion towards the summits of hills. The sacred status of a site was not always related to its high altitude, like those that were used for rites of sacrifice (*2*), but also for presenting special geographical and geological characteristics (*3*). These revered hills were part of the religious life and were the scene of intense ritual practices (*4*).

For the Incas, as they expand their empire, it was natural to appropriate the sites of worship of local cultures, and then give them a new religious meaning. The activity of re-signifying a cult site used to be part of the ritual conquest strategies implemented by the Incas throughout the Tawantinsuyu (*5*). Resent archeological findings supports the hypothesis that there existed, prior to the arrival of the Spaniards, an important Inca occupation in the Central Valley of Chile, and it replicated characteristics related to the sacred geography of the Valley of Cuzco with its own ceque system based on astronomical observations (*6–10*).

It could be argued that within the archaeological sites allocated to the Inca culture in Chile, one of the most important and enigmatic is found in Chada Valley, located in the province of Maipo, Metropolitan Region (Figure 1A).

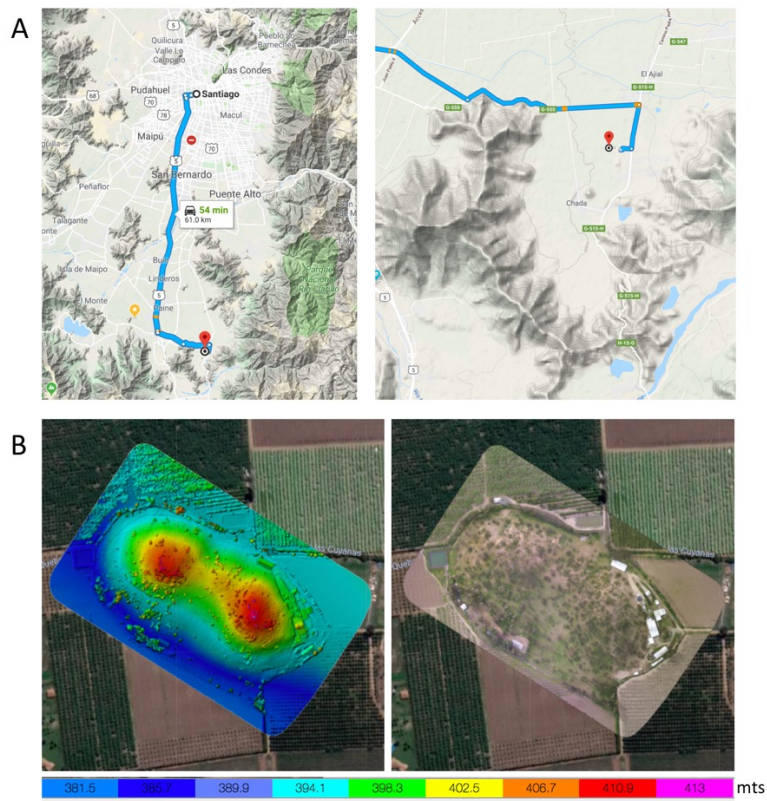

Figure 1

The meaning of the Ruins of Chada represents one of the greatest mysteries regarding Inca occupation in Chile. Various theories have been proposed regarding function and meaning for the complex architecture present in this site. Initially, as with most of the Inca buildings, it was given the role of "Pukara" or fortress (*11*). Then, it was proposed that it could function as a "space organizer" to annex the southern lands of the Collasuyu (*12*). Later, it was proposed that it could be considered as an organizer for imaginary lines or ceques originating from the center of the ruins which point at different places of importance for the Inca. The directions would be given by alignments of astronomical events such as solstices and equinoxes in conjunction with natural horizon markers (*13*). However, none of these theories have significantly explained the function and symbolism concealed in this shrine´s blueprint. The layout of the foundations and the complex design present in this site has no equivalent to any other Inca site known in Chile.

**Results**

In September 2016 we made a high-resolution mapping of the hill where the ruins of Chada are located. The mapping consisted in taking 265 georeferenced photographs over an altitude of approximately 40 meters over the summit. This map covers an area of 9.55 hectares, enclosing the entire hill of Chada ruins. Then, by image processing we obtained a georeferenced orthophoto with size 16339 x 14753, and resolution 2.6 cm/pixel, in conjunction with an elevation map (Figure 1B). Based on the existing planimetry of the site (*11*), we readjusted the orientation and shape of the wall´s layout, obtaining an updated and georeferenced map (Figure 2). The central well or ushnu was positioned in the exact center with respect to the walls. This point was not built above the ground like the rest of the walls, but was dug into the rock of the hill (Figure S1).

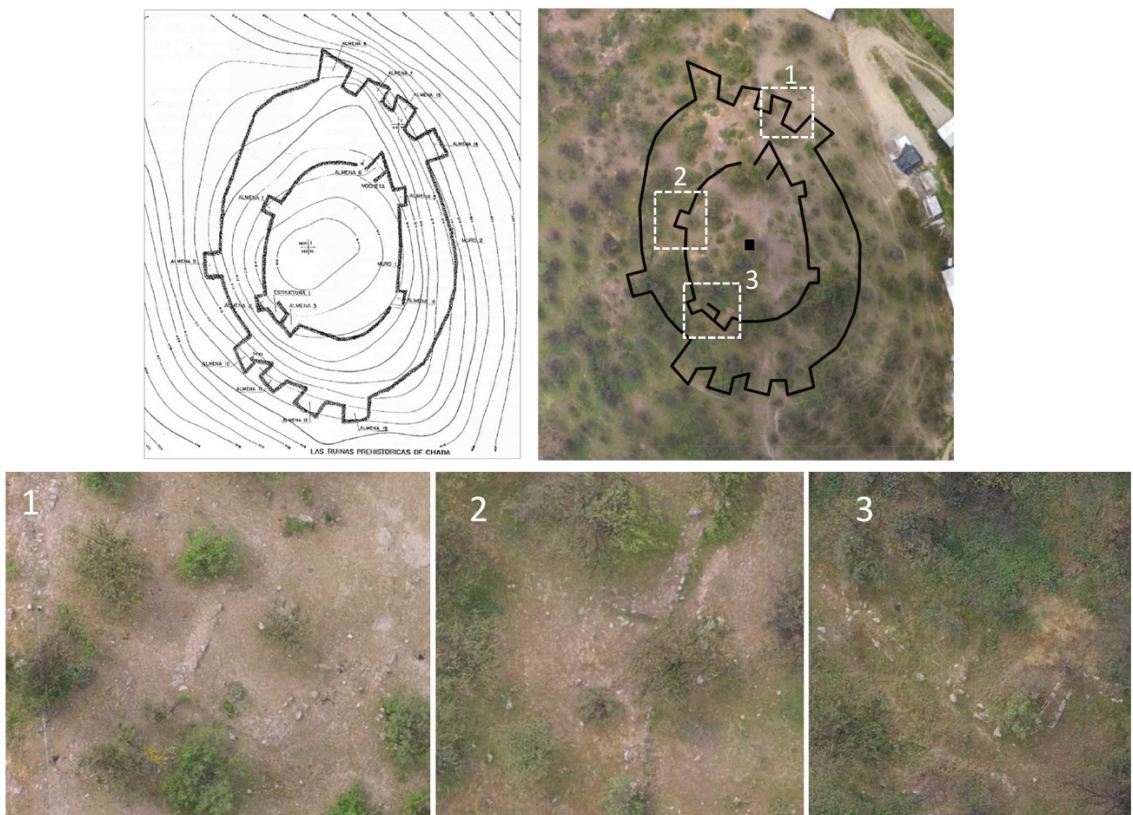

Figure 2

The outline of the walls implicitly expresses a significant symbolism. To try to understand the meaning behind this architecture we deconstructed the basic signs forming the structure (Figure 3). We can differentiate four distinctive basic elements: a center or origin; concentric circles; rectangular "outbounds" or salients in opposite positions relative to the center; and finally two signs or symbols which can be interpreted as mountain peaks in opposite positions with respect to the center. This symbol could represent one or two peaks of large hills, and the opposite symbol could represent one or two summits of smaller hills. Also in the symbol referring to the small hill we can see a lower line that could represent the arrival of a road or the presence of walls.

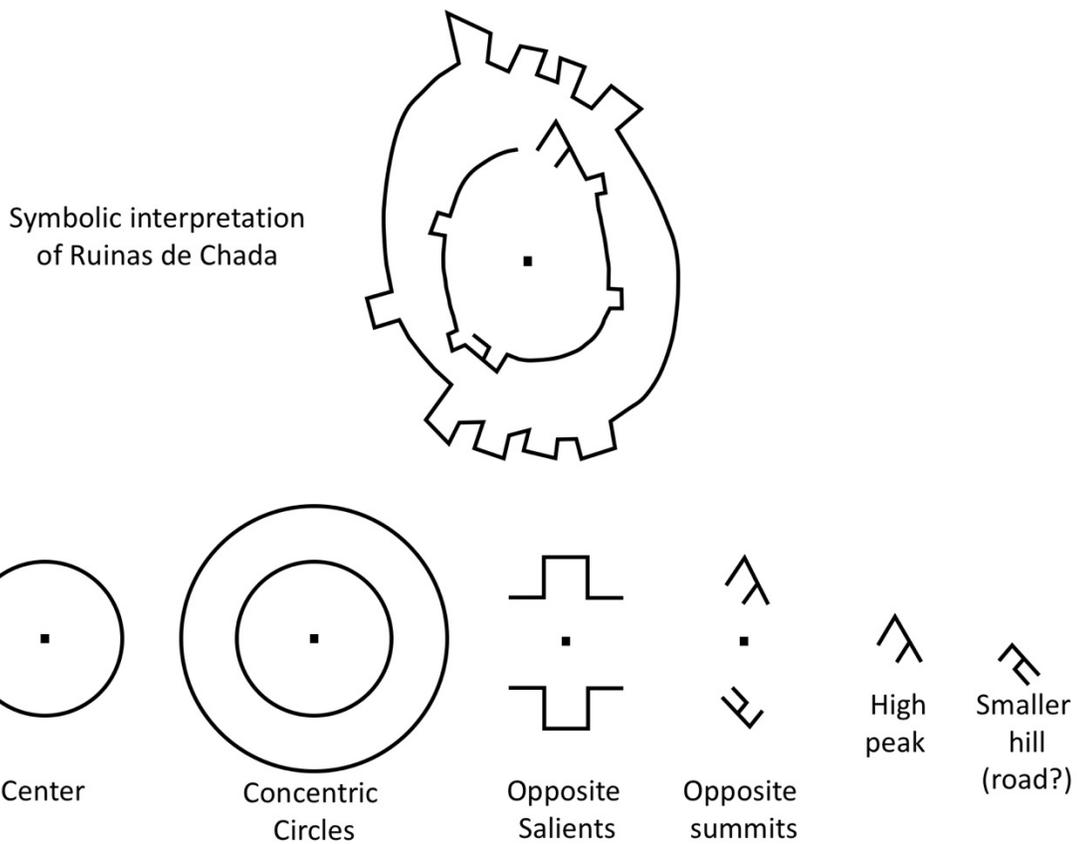

Figure 3

Based on this interpretation we made an extensive search on Google Earth of the surrounding hills. We discovered that each pair of opposing salients, originally described as "battlements" (*11*), points toward the top of a large hill or volcano in the Andes mountain range, and on the opposite side toward the summit of smaller hills in which other Inca ruins were built (Figure 4C). We color-coded and named each ceque line from 1st to 7th, being the first and seventh ceques the ones that point towards hills in the valley nearer and farther from the Ushnu of Chada, respectively (Figure 4A and S2-S4).

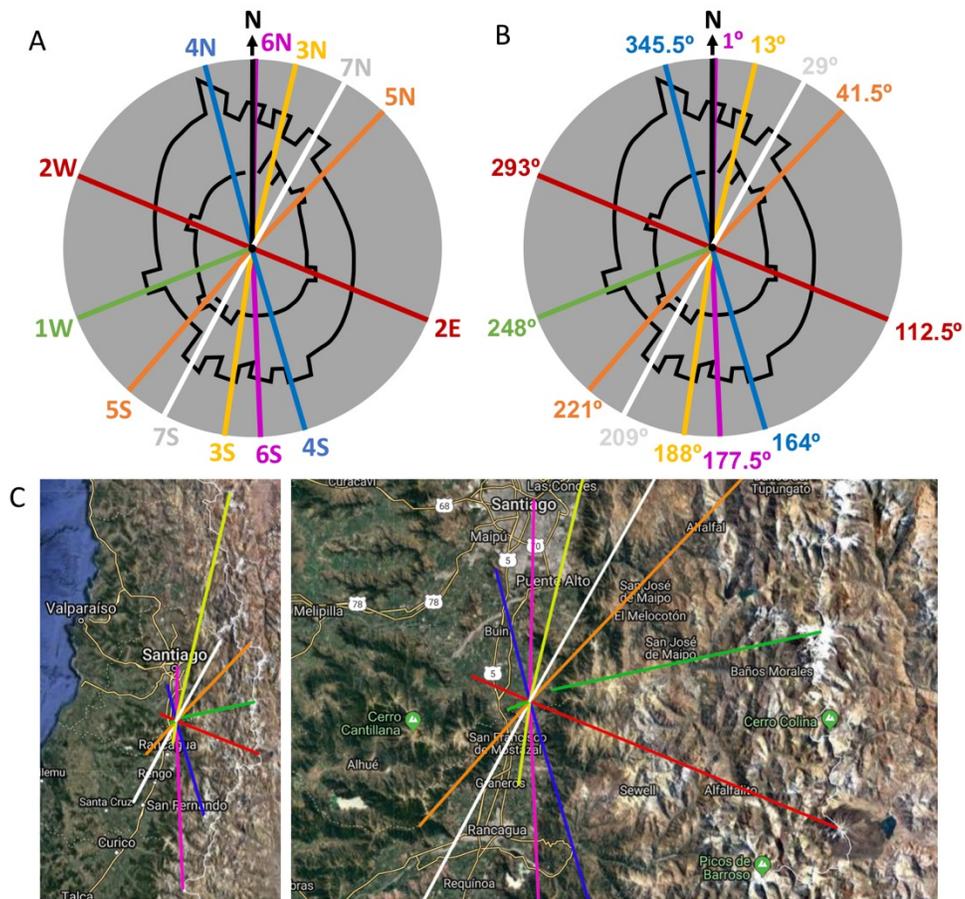

Figure 4

### 7th ceque (Grey)

Towards North (7N) it passes through salient number 14 of the lower wall, originally listed as battlement number 14 (*11*), and goes to the top of Cerro El Plomo (Figure S2A), a well-known sacred mountain for the Incas and in which an Inca Capacocha rite was performed. Towards South (7S) it passes through salient number 10 of the lower wall and goes to the highest peak within the chain of mountains called Yaquil Range (Figure S2A),, which borders the old Tagua-Tagua Lagoon. In the northern part of this chain is Cerro La Muralla, where other Incas ruins are located, being the most southerly known Inca structures of the empire (*14, 15*).

### 6th ceque (Fuchsia)

Towards south (6S) it passes through salient number 12 of the lower wall and goes to the volcanic complex Planchón-Peteroa (Figure S2B). The final point of the Ceque was chosen as the top of volcano Azufre I as it is the highest of the four volcanoes of the complex. It is to be noted, however, that the four summits are relatively aligned from north to south following the direction of the Ceque. Towards North (6N) the Ceque passes through salient number 7 and goes to the top of

Cerro Santa Lucia in the center of Santiago (Figure S2B). Cerro Santa Lucia is known as an Inca Huaca related to the astronomical cult, in which there are still sacred milestones. i.e. stone carved as a staircase (*8*, *9*).

**5th ceque (Orange)**

Towards East (5E) it passes through salient number 5 of the upper wall and goes to the top of Cerro Polleras (Figure S2C). Towards west (5W) it passes through salient number 2 of the upper wall and goes to the top of the hill next to Cerro Trentrén, in Doñihue (Figure S2C). Cerro Trentrén is located on the banks of Cachapoal river and it is known for being a sacred hill for the Mapuche community and Diaguita-Inca groups. An Inca shrine is located over this hill (*16*).

**4th ceque (Blue)**

Towards South (4S) it passes through salient number 13 of the lower wall and goes to the top of Cerro Alto de los Arrieros in Sierra del Brujo (Figure S3A). Towards North (4N) it passes through salient number 8 of the lower wall and goes to the highest peak of Cerro Chena (Figure S3A). On a subpeak of Cerro Chena is located the Pukara or Huaca of Chena, one of the most studied Inca site in Chile (*6*, *17*, *18*).

**3rd ceque (Yellow)**

Towards North (3N) it passes through salient number 15 of the lower wall and goes to the top of Cerro Mercedario (Figure S3B), one of the highest peaks in South America, and in which an Inca Capacocha rite was performed (*1*). Towards South (3S) it passes through salient number 11 of the lower wall and goes to the top of Cerro Grande de la Compañia (Figure S3B). At the top of this hill there is also a "Pukara" (*19*).

**2nd ceque (Red)**

Towards East (2E) it passes through salient number 4 of the upper wall and goes towards the top of Maipo Volcano (Figure S3C). Towards West (2W) it passes through salient number 1 of the upper wall and goes towards the top of Cerro Cullipeumo (Figure S3C). This hill also presents on its summit walls attributed to the Inca occupation (*17*, *20*).

**1st ceque (Green)**

The first ceque is complex, since in the organizer of Chada only one salient is pointing to West, without its respective opposite to East. Towards West (1W) it passes through salient number 9 of the lower wall and heads towards the top of Cerro Challay (Figure S4A). Cerro Challay, formerly Tallay, is also a well-known Inca sacred hill (*12*). If we extend the ceque to the East or opposite direction (1E´) we did not reach any significant high peak. However, we reached the summit of Cerro El Peral, where there is an Inca shrine (*21*). By reviewing the shape of the ruins of Cerro El Peral we realized that it has some similarity with Chada, where there is a circular wall in which there is only one salient pointing to East. As if the opposite salient to number 9 missing in the ruins of Chada would have been built especially at the top of Cerro El Peral. If we extend a line from the center and pointing towards this salient (1E´´) we find that it goes straight to the top of Cerro Marmolejo (Figure S4B). In this way, the first Ceque would connect Chada with the top of Cerro Challay, and the top of Cerro Marmolejo with Cerro el Peral to follow the pattern of high and small mounts.

**Northern Ceque (Black)**

Within the upper wall there is an opening on the north side that would indicate an entrance towards the Ushnu. However, there is no similar entry in the lower wall, hence we believe that this opening does not mean a door or entrance for people's access, but it also has a symbolic meaning and directionality regarding the location of some important place. By extending a line from the ushnu to the geographical North passing through the center of the opening (arrow N) We arrived exactly at the center of the Santiago Metropolitan Cathedral, next to Plaza de Armas (Figure S5). Some archeological studies have shown that an important Inca temple is located beneath the cathedral floor (*22*). Therefore, the north gate of the upper wall might signify a direct connection between the Ruins of Chada and the ancient Inca temple located under the current Metropolitan Cathedral.

**Alignment Accuracy**

These alignments have a significant level of precision. When calculating the degrees of the angles formed with respect to the geographical north or Northern Ceque (Figure 4B), we can determine the deviation that each ceque has when passing through the Ushnu (difference in degrees from 180°: seventh ceque 0°; sixth ceque 3.5°; fifth ceque 0.5°; fourth ceque 1.5°, third ceque 5°; second ceque 0.5°). Being the third ceque the one with highest degree of deviation, but in turn the one that covers the most extensive distances.

**Sacred Geometry Resulting from Alignments**

To analyze the meaning behind the encoded alignments in Ruins of Chada, a table was constructed with GPS coordinates and distances in kilometers between the ushnu of Chada and the peaks of all indicated mounts (Table S1). Then, following our symbolic interpretation, we built a series of concentric rings with center of origin in the Ushnu of Chada and extending the radius to all the shrines or "Pukaras" in each one of the small hills indicated by the ceques (Figure S6). The distance between the Ushnu and the shrine locations was chosen as radius of the circles (Figure 5). In the case of the first Ceque, the chosen radius was the distance between Chada and Cerro Challay (Figure S7A).

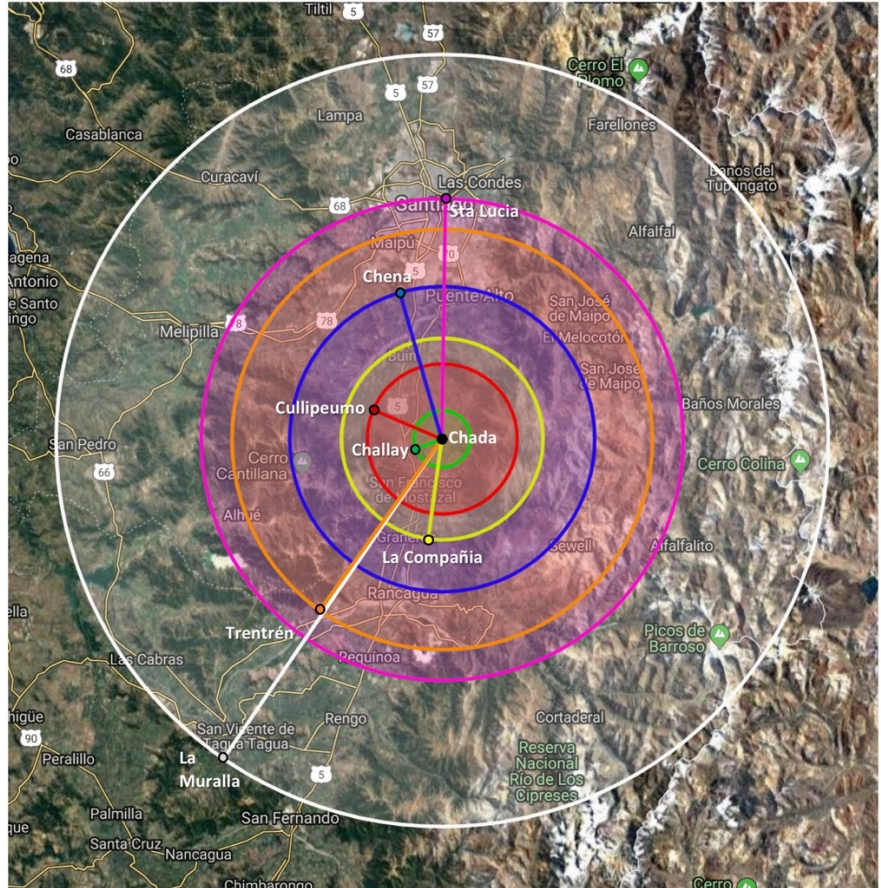

Figure 5

By positioning the seven concentric circles, colored in the same way as the ceques, we found a setup with a progressive and sequential ordering (Figure 5). The shrines alternate between North-South axis as they approach the center. In addition, the relationship between the length of the diameters has, at first glance, a certain proportion (Figure 6B). For this reason, we built a grid where the lengths of the diameters of each circle are represented in order to study their respective proportion (Figure 6C).

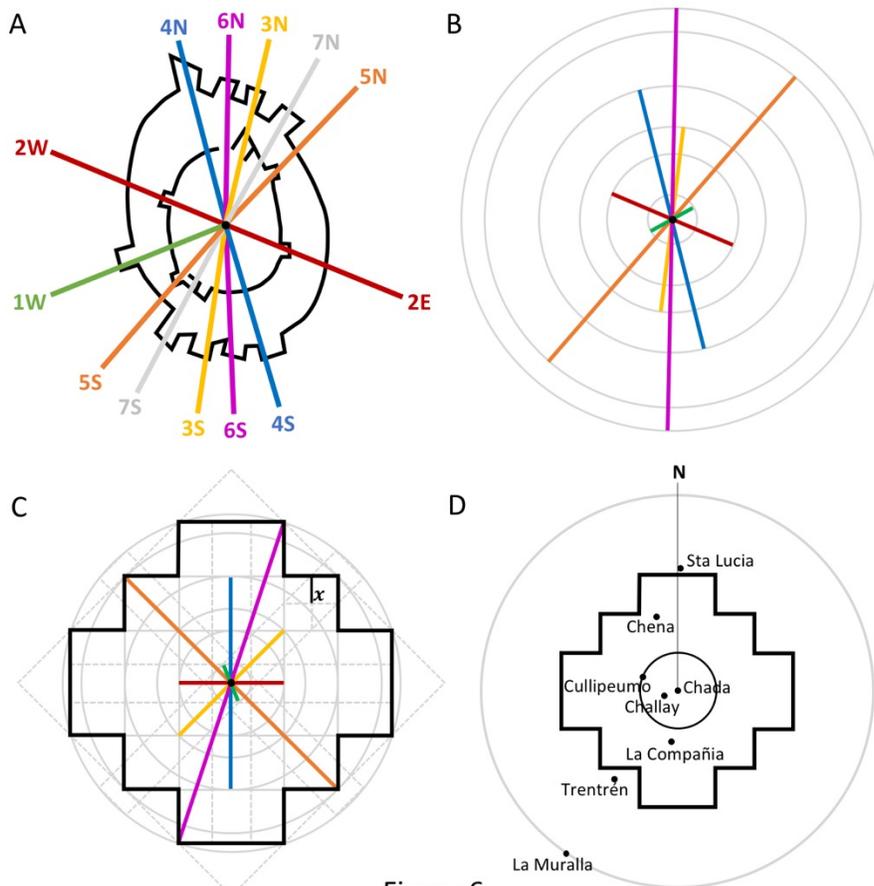

Figure 6

The relationship between the circle diameters is similar to the different proportions that emerge when building the Chakana symbol (Figure 6C-D). This symbol is one of the most recurrent and sacred in all native Andean cultures (*23*), and is also the geometric form by which the geography of Cuzco and Tihuanacu valleys were conceived and sacralized (*24*). To calculate the level of precision in which the concentric circles generated by the shrine positions form the geometric Chakana symbol, we built a set of simple equations that determine all the diameters necessary to form this symbol based on a known distance $x$, equivalent to the length of the side of the smaller square in the grid (Figure 6C and Table S2). Taking as a measure $x = 7,72$ km, equivalent to half of the length between the Ushnu of Chada and the top of Cerro Cullipeumo, we can generate a set of calculated diameters to compare with "existing" diameters (Table S2). All proportions are found within an accuracy of 95% (calculated error: 6$^{th}$ ceque 1.48%; 5$^{th}$ ceque 0.85%; 4$^{th}$ ceque 1.36%, 3$^{rd}$ ceque 5.19%; 1$^{st}$ ceque 4.39%).

**Discussion and Conclusion**

The 2$^{nd}$ and 5$^{th}$ ceques are generated by the same upper wall, which puts them into the same category. We propose that Cerro Cullipeumo y Cerro Trentrén fall into the category of Pakarinas, or places of origin, and it is likely that these hills were sacralized previously by local cultures prior to the arrival of the Incas. Even though the 7th ceque does not participate in the formation of the Chakana symbol we believe this circle represents the outer limit of the state occupation (Figure 7).

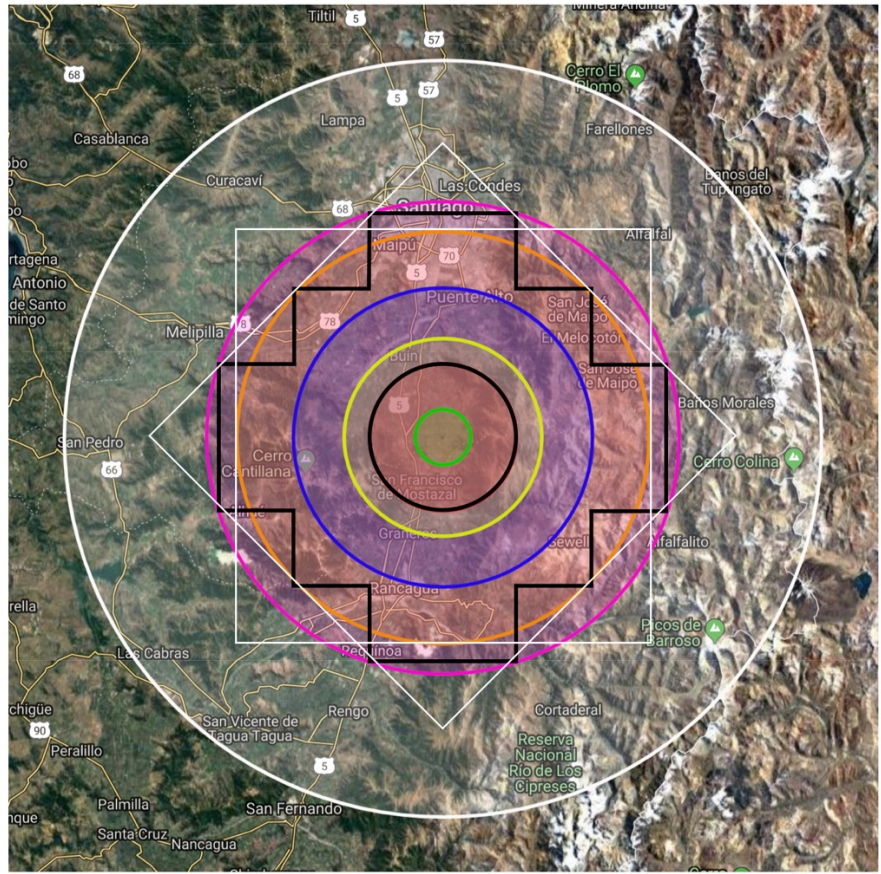

Figure 7

The emergence of the Chakana symbol from the position pattern of the identified shrines supports the notion that the precise location and interrelation of these sacred hills were deliberately found and sacralized by the Incas, in a way similar to the position of sanctuaries and temples of Peru (*23, 24*). Somehow, Inca astronomers were able to perform precise geographic measurements in order to design this "spatial organizer". The Chakana symbol represents part of the worldview of the native people of the Andes and is derived from sacred proportions or Tupus found among the stars that form the Southern Cross. Through this work of sacralization the Incas might wanted to "bring heaven to earth".

These findings suggest that other Inca shrines throughout south America might also possessed encrypted information in their structures and wall layouts. We hope these findings might help to encourage the development of better government policies regarding preservation and conservation of archeological sites.

# References

1. C. Vitry, Caminos Rituales y Montañas Sagradas. Estudio de la Vialidad Inka en el Nevado de Chañi, Argentina. *Boletín del Mus. Chil. Arte Precolomb.* **12**, 69–84 (2007).
2. P. Duviols, La Capacocha. Mecanismo y Función del Sacrificio Humano, su Proyección geométrica, su papel en la política integracionista y en la economía redistributiva del tawantinsuyu. *Allpanchis*, 11–57 (1976).
3. B. S. Bauer, *The Sacred Lansdcape of the Inca: The Cuzco Ceque System* (University of Texas Press, Austin, 1998).
4. J. Berenguer, *Unkus ajedrezados en el arte rupestre del sur del Tawantinsuyu: ¿La estrecha camiseta de la nueva servidumbre?* (2013).
5. P. Cruz, Huacas olvidadas y cerros santos. Apuntes metodológicos sobre la cartografía sagrada en los Andes del sur de Bolivia. *Estud. Atacameños*. **38**, 55–74 (2009).
6. P. Bustamante, La Huaca del Cerro Chena. Arquitectura Sagrada del Pueblo Inca. *Rev. CIMIN (Construcción Ind. y Minería*, 1–6 (1996).
7. P. Bustamante, Santiago del Nuevo Extremo ¿Una Ciudad Sin Pasado? *Diseño Urbano y Paisaje (Universidad Cent. Chile)* (2006).
8. P. Bustamante, R. Moyano, Cerro Wanguelen: obras rupestres, observatorio astronómico-orográfico Mapuche-Inca y el sistema de ceques de la cuenca de Santiago. *Rupestreweb* (2013).
9. A. López Tapia, La Sagrada Función del Cerro Santa Lucía y la Fundación de Santiago. *Rev. Chil. Hist. y Geogr. Sección Geogr.* (2013).
10. R. Stehberg, G. Sotomayor, Mapocho Incaico. *Boletín del Mus. Hist. Nat. Chile*. **61**, 85–149 (2012).
11. M. T. Planella, R. Stehberg, Intervención Inka en un territorio de la cultura local Aconcagua de la zona centro-sur de Chile. *Tawantinsuyu*, 58–78 (1997).
12. M. C. Odone, El Valle de Chada: La Construcción Colonial de un Espacio Indígena de Chile Central. *Historia Santiago*. **30**, 189–209 (1997).
13. N. Ruano, Arqueoastronomía Inca en el Sitio Ruinas de Chada, RM, Chile. *Actas del XIX Congr. Nac. Arqueol. Chil.*, 133–140 (2012).
14. R. Stehberg, Fortaleza "La Muralla" (Laguna de Tagua-Tagua). *Moticiario Mens. Mus. Nac. Hist. Nat.* **XIX**, 3–6 (1974).
15. J. Sepúlveda, A. San Francisco Araya, B. Jimenez Belvar, S. Perez Lizana, *El Pucara del Cerro La Muralla: Mapuches, Incas y Españoles en el Valle del Cachapoal* (2014).
16. R. Stehberg, A. Rodriguez, Ofrendas Mapuche-Incaicas en el Cerro Tren Tren de Doñihue. *Tawantinsuyu*, 29–35 (1995).
17. R. Stehberg, La Fortaleza de Chena y su relación con la Ocupación Incaica de Chile Central. *Publ. Ocas. Mus. Nac. Hist. Nat.*, 3–37 (1976).
18. R. Stehberg, G. Sotomayor, C. Gatica, El Paisaje Ritualizado del Pucará de Chena. *XIX Congr. Arqueol. Paisaje, Astron. y Ritual. en Los Andes Cent. Sur*, 141–147 (2012).

19. M. T. Planella, R. Stehberg, B. Tagle, H. Niemeyer, C. Del Rio, La Fortaleza Indígena del Cerro Grande de la Compañía y su Relación con el Proceso Expansivo Meridional Incaico. *Actas del XII Congr. Nac. Arqueol. Chil.*, 403–421 (1993).

20. A. Troncoso, "Conservacion y Difusión Pucara Cerro Collipeumo, Región Metropolitana y Trabajos Anexos en Cerro Chena" (2010).

21. R. Stehberg, Caminos, Guacas y el Reducto Fortificado de Cerro el Peral: Instalaciones Para el Control Inca del Paso de Chada, Chile Central. *Boletín del Mus. Hist. Nat. Chile*, 129–146 (2013).

22. R. Stehberg, C. Prado, P. Rivas, El Sustrato Incaico de la Catedral Metropolitana. *Boletín del Mus. Hist. Nat. Chile*. **66**, 161–208 (2017).

23. C. Milla V., *Ayni: Semiótica Andina de los Espacios Sagrados* (Universidad particular San Martín de Porres, 1. ed., 2003).

24. M. Scholten, *La Ruta de Virakocha* (1985).

# Supplementary Material

## Materials and Methods

To generate an orthophoto of the site we used a DJI Phantom 3 drone equipped with a DJI FC300S camera. Automatic mapping and photo acquisition was performed using the Map Pilot application from Maps Made Easy (https://www.mapsmadeeasy.com). Alignments, circles and distances were calculated using the Maps Made Easy application. We provide public access to this map in order for the scientific community to review the alignments and study further the size and position of the shrine walls (www.mapsmadeeasy.com/maps/public/7a9761660c2c490e9eb8cfcf93a142b8). Peak elevation and name of mountains were found in at military geographic institute of Chile (www.igm.cl). Altitude of smaller hills and GPS coordinates were found using google earth (www.google.com/earth).

## Supplementary Figures and Tables

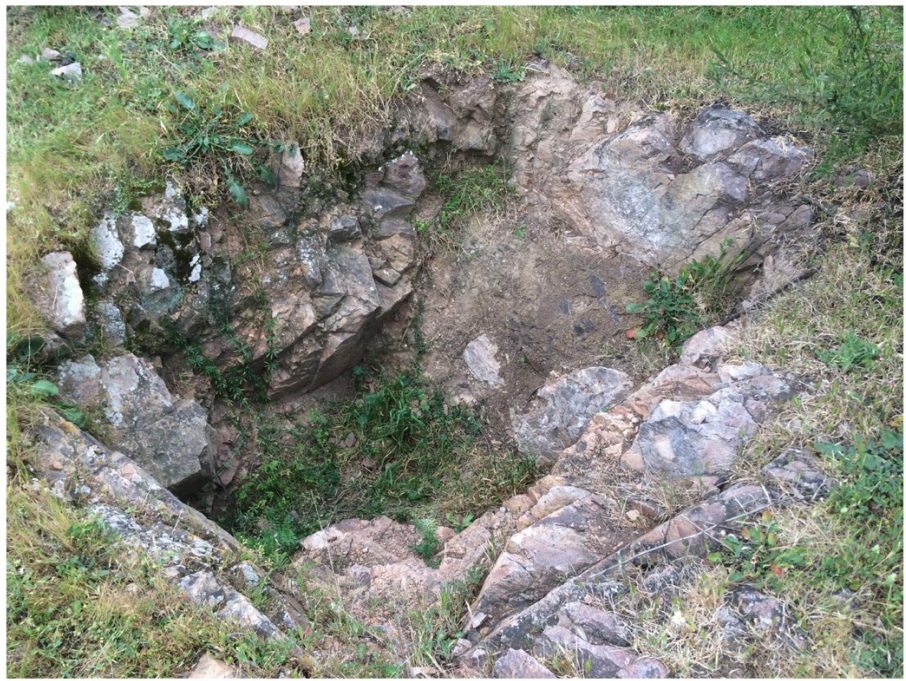

Figure S1

A 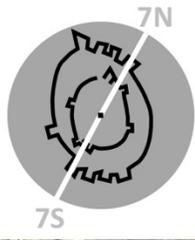  B 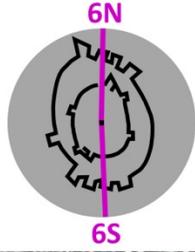  C 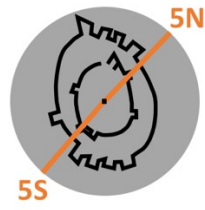

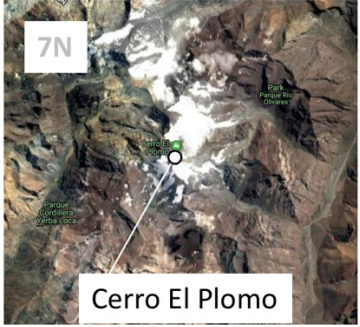
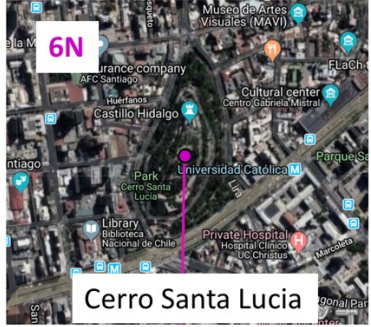
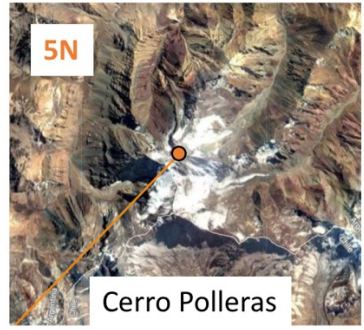

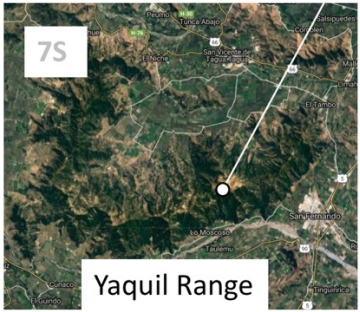
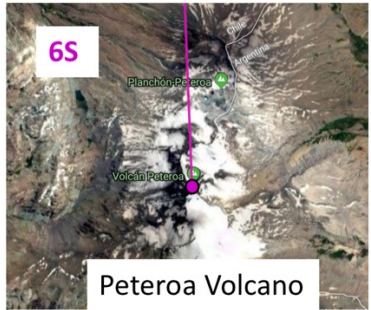
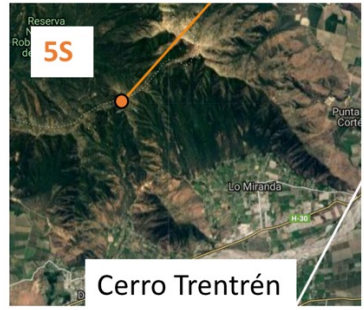

Figure S2

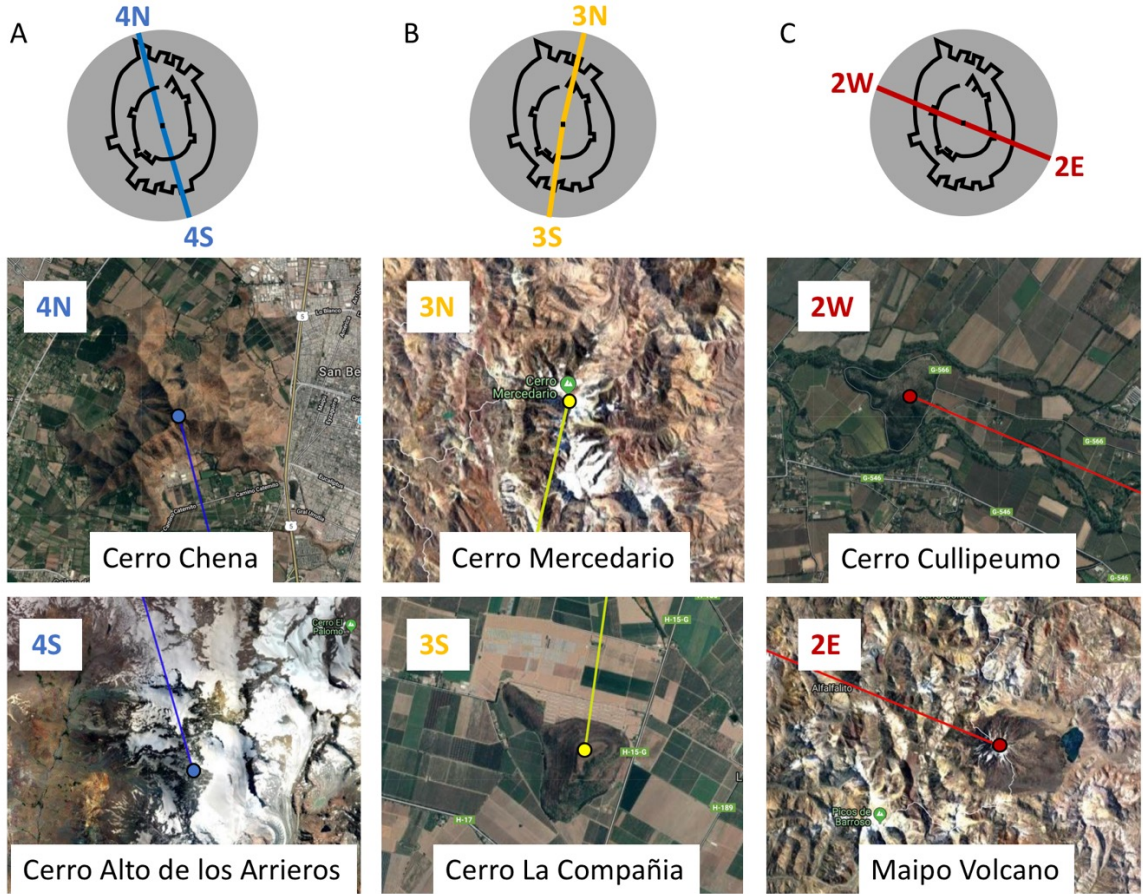

Figure S3

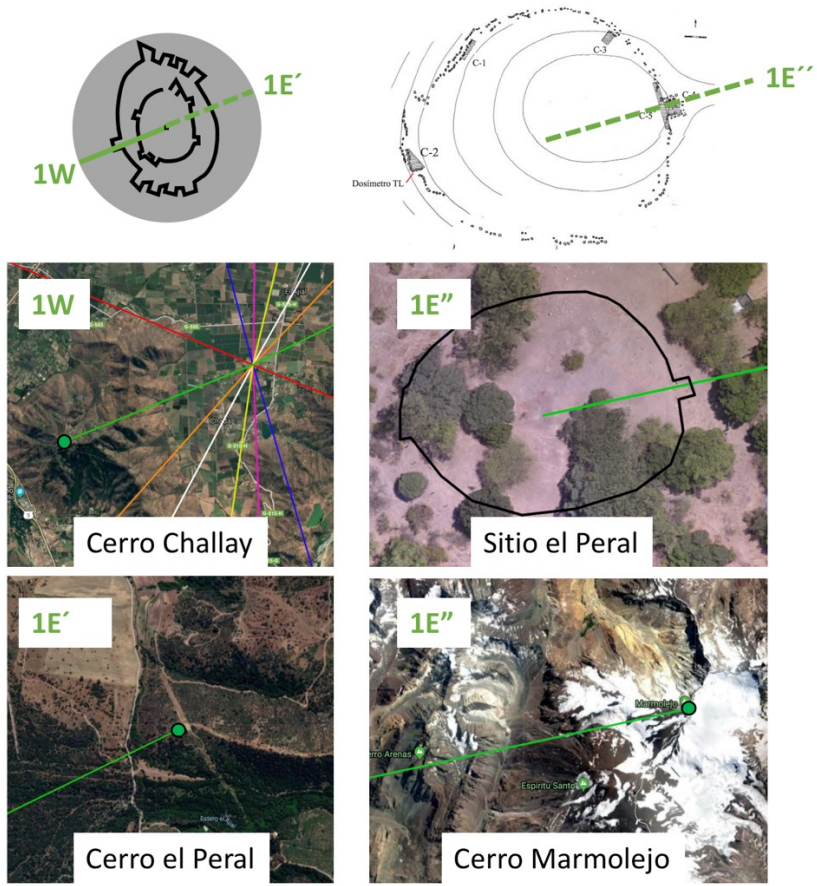

Figure S4

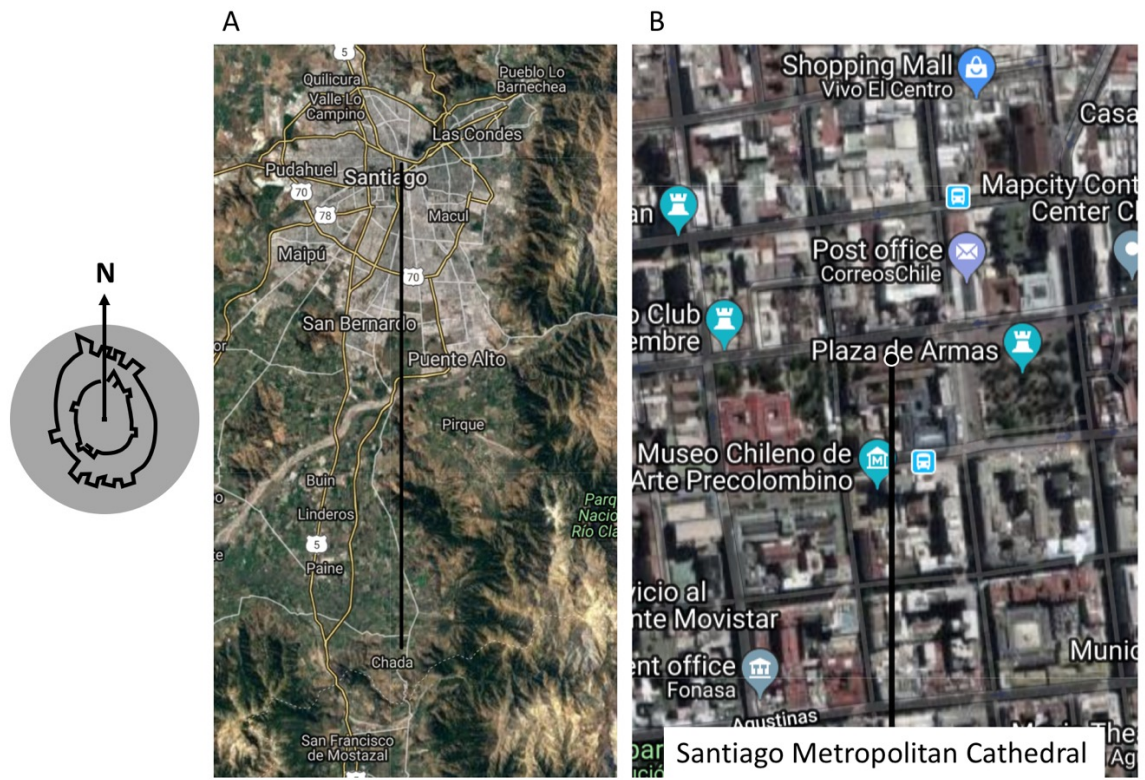

Figure S5

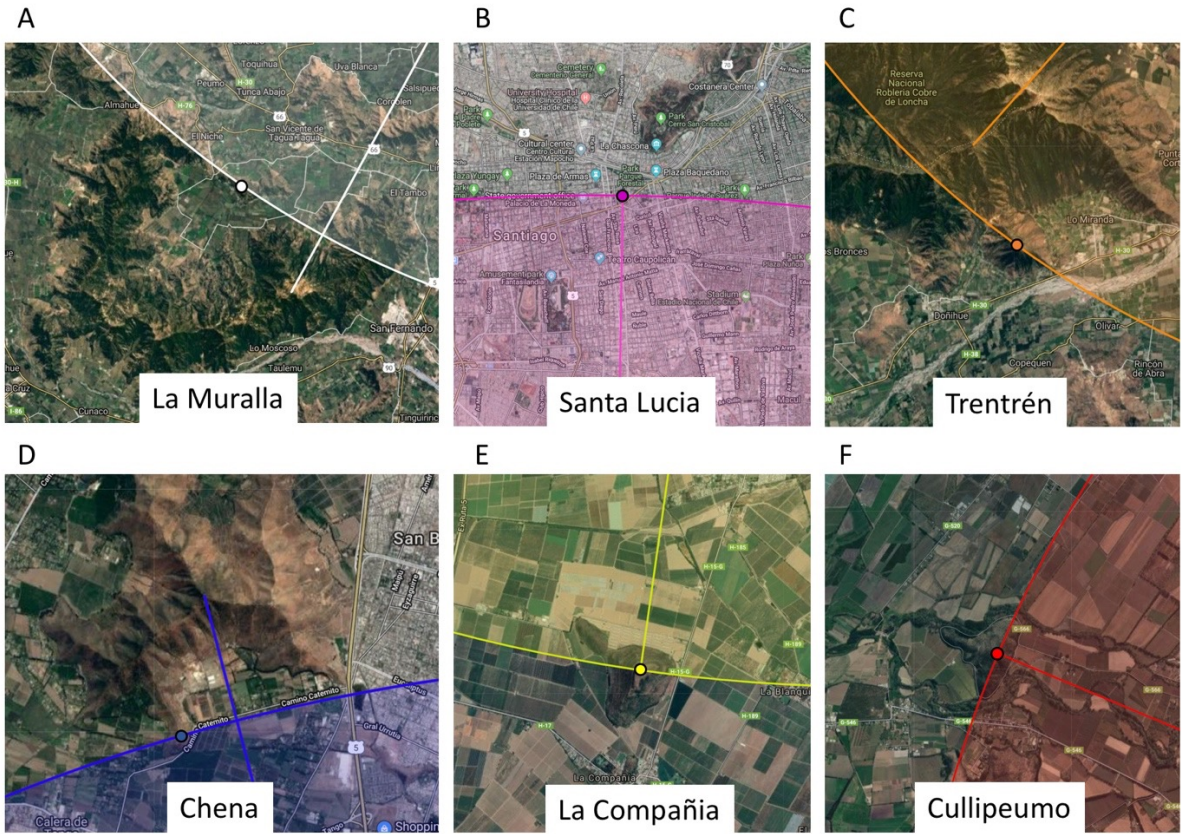

Figure S6

A 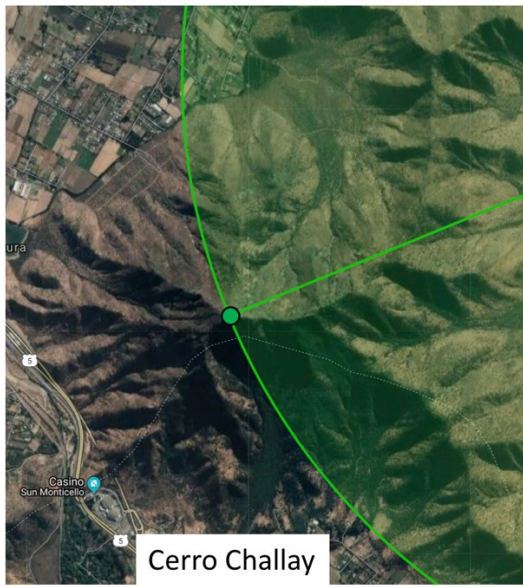 B 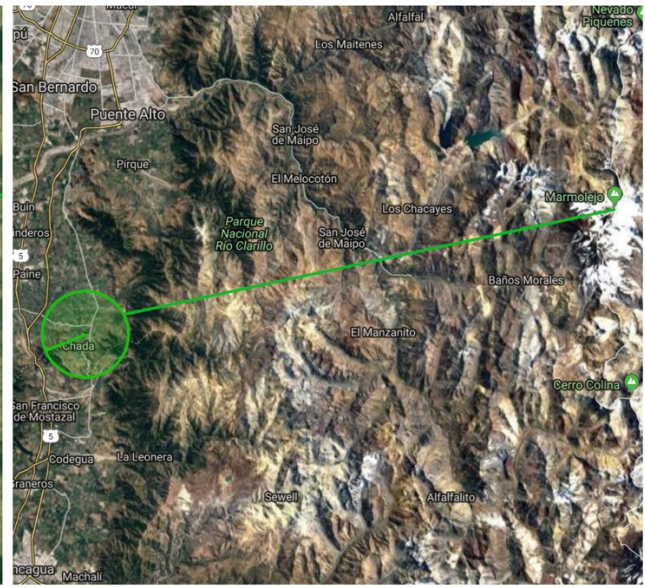

Cerro Challay

Figure S7

| Hill/Mountain/Volcano/Place | Elevation (masl) | Coordinates | Distance from Chada (Km) |
|---|---|---|---|
| Chada | 451 | 33°53'9.63"S;  70°39'6.84"W | 0 |
| Challay | 2280 | 33°54'21.22"S; 70°42'36.31"W | 5.7 |
| El Peral | 710 | 33°51'41.31"S; 70°35'30.14"W | 6.2 |
| Cullipeumo | 503 | 33°49'54.27"S; 70°48'21.00"W | 15.4 |
| Grande de La Compañía | 677 | 34° 4'5.83"S;  70°40'59.95"W | 20.7 |
| Chena Pukara | 625 | 33°36'54.44"S; 70°44'48.70"W | 31.3 |
| Chena (highest peack) | 932 | 33°35'38.91"S; 70°44'34.72"W | 33.5 |
| Trentrén (Norwest peak) | 1820 | 34° 9'40.87"S; 70°56'51.39"W | 40.9 |
| Trentrén | 960 | 34°12'9.12"S;  70°55'31.48"W | 43.2 |
| Santa Lucia | 629 | 33°26'25.15"S; 70°38'36.53"W | 49.5 |
| Metropolitan Cathedral of Santiago | 578 | 33°26'15.65"S;  70°39'6.84"W | 49.8 |
| Marmolejo | 6108 | 33°44'3.09"S;  69°52'42.97"W | 67.4* |
| La Muralla | 458 | 34°28'28.02"S;  71° 8'28.40"W | 79.6 |
| V. Maipo | 5264 | 34° 9'44.65"S; 69°50'13.40"W | 81.5 |
| El Plomo | 5424 | 33°14'10.45"S; 70°12'51.94"W | 82.9 |
| Chain of Yaquil | 1172 | 34°33'29.37"S;  71° 5'31.52"W | 85 |
| Alto de los Arrieros | 4990 | 34°39'37.83"S; 70°22'44.76"W | 89.7 |
| V. Azufre I | 4113 | 35°16'22.85"S; 70°34'54.01"W | 154.3 |
| Mercedario | 6770 | 31°58'44.25"S; 70° 6'45.29"W | 217.9 |
| * Distance from El Peral | | | |

Table S1

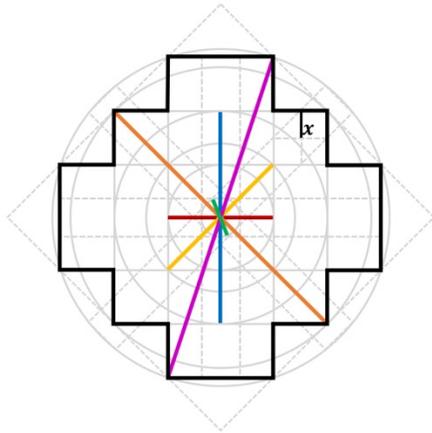

| Formula | Calculated (Km) | Existing (Km) | Error % | Mount |
|---|---|---|---|---|
| $x\sqrt{2}$ | 10.92 | 11.4 | 4.39 | Challay |
| $4x$ | 30.88 | 30.88 | - | Cullipeumo |
| $4x\sqrt{2}$ | 43.67 | 41.4 | 5.19 | La Compañia |
| $8x$ | 61.76 | 62.6 | 1.36 | Chena |
| $8x\sqrt{2}$ | 87.34 | 86.6 | 0.85 | Trentrén |
| $4x\sqrt{10}$ | 97.65 | 99.1 | 1.48 | Santa Lucia |

$x = 7.72 \text{ km}$

Table S2